\documentclass[useAMS,usenatbib]{mn2e}
\usepackage{multicol}
\usepackage{multirow}
\usepackage{natbib}

\usepackage{graphicx}
\usepackage{txfonts}
\usepackage{psfig}
\usepackage{subfigure}
\usepackage{natbib}
\title[Clumps and triggered star formation in ionised molecular clouds]
{Clumps and triggered star formation in ionised molecular clouds}
\author[Walch et al.]
{S. Walch$^{1}$\thanks{E-mail: walch@mpa-garching.mpg.de}, 
A.P.~Whitworth$^{2}$, T.G.~Bisbas$^{3}$, R.~W\"unsch$^{4}$, D.A.~Hubber$^{5,6}$\\
$^{1}$Max-Planck-Institute for Astrophysics, Karl-Schwarzschild-Str. 1, 85741 Garching, Germany \\
$^{2}$School of Physics \& Astronomy, Cardiff University, 5 The Parade, Cardiff CF24 3AA, Wales, UK \\
$^{3}$Department of Physics \& Astronomy, University College London, Gower Place, London WC1E 6BT, UK\\
$^{4}$Astronomical Institute, Academy of Sciences of the Czech Republic, Bocni II 1401, 141 31 Prague, Czech Republic\\
$^{5}$Technical University Munich, Excellence Cluster Universe, Boltzmannstr. 2, 85748 Garching, Germany\\
$^{6}$University Observatory Munich, Department of Physics, Ludwig-Maximilians-University Munich, Scheinerstr.1, 81679 Munich, Germany. }

\begin{document}

\date{Accepted . Received 2012 June 27; in original form }

\pagerange{\pageref{firstpage}--\pageref{lastpage}} \pubyear{2012}

\maketitle

\label{firstpage}

\begin{abstract}
Infrared shells and bubbles are ubiquitous in the Galaxy and can generally be associated with H{\sc ii} regions formed around young, massive stars. In this paper, we use high-resolution 3D SPH simulations to explore the effect of a single O7 star emitting photons at $10^{49}\,{\rm s}^{-1}$ and located at the centre of a molecular cloud with mass $10^4\,{\rm M}_{_\odot}$ and radius $6.4\,{\rm pc}$; the internal structure of the cloud is characterised by its fractal dimension, ${\cal D}$ (with $2.0\leq {\cal D}\leq 2.8$), and the variance of its (log-normal) density distribution, $\sigma_{_{\rm O}}^2$ (with $0.36\!\leq\!\sigma_{_{\rm O}}^2\!\leq1.42$). Our study focuses on the morphology of the swept-up cold gas and the distribution and statistics of the resulting star formation. If the fractal dimension is low, the border of the H{\sc ii} region is dominated by extended shell-like structures, and these break up into a small number of massive high-density clumps which then spawn star clusters; star formation occurs relatively quickly, and delivers somewhat higher stellar masses. Conversely, if the fractal dimension is high, the border of the H{\sc ii} region is dominated by a large number of pillars and cometary globules, which contain compact dense clumps and tend to spawn single stars or individual multiple systems; star formation occurs later, the stellar masses are somewhat lower, and the stars are more widely distributed.
\end{abstract}


\section{Introduction}%

Infrared shells and bubbles are ubiquitous in the Galaxy and can often be associated with H{\sc ii} regions around young, massive stars \citep{Deharveng2010, Simpson2012}, which emit ionising radiation having $E_\gamma > 13.6\; {\rm eV}$. The ionised gas is heated to $\,\sim\! 10^4\,{\rm K}$, and the resulting pressure increase causes the H{\sc ii} region to expand, sweeping up and compressing the much colder ($\sim 10\;{\rm to}\;30\,{\rm K}$) surrounding molecular cloud. Since molecular clouds typically have a complicated, clumpy internal structure, the ionising radiation penetrates to different depths in different directions, producing highly irregular ionisation fronts. Thus, evolved H{\sc ii} regions have diverse morphologies, sometimes appearing as perfectly round shells like RCW 120 \citep{Deharveng2009}, sometimes filamentary and/or clumpy with large holes through which the ionising radiation can escape, like Carina \citep{Smith2010}. 

In \citet[][hereafter W12]{Walch2012} we show that all of these morphological features can be reproduced by invoking different values for the fractal dimension, $\mathcal{D}$, of the molecular cloud into which the H{\sc ii} region expands. Low $\mathcal{D}$ (i.e. $\mathcal{D} \le 2.2$) corresponds to clumpy clouds in which the density sub-structure is dominated by large-scale fluctuations. As $\mathcal{D}$ is increased, small-scale fluctuations become increasingly more important. As a result, W12 report a morphological transition from {\it shell-dominated} H{\sc ii} regions for low $\mathcal{D}$, to {\it pillar-dominated} H{\sc ii} regions for large $\mathcal{D}$. In this paper we investigate the formation of cold clumps at the boundaries of H{\sc ii} regions, and the triggering of star formation in them. We show that the statistical properties of the cold clumps, and of the stars that they spawn, are both correlated with the fractal dimension of the initial molecular cloud. In this context, two distinct modes of triggered star formation have traditionally been defined and contrasted: {\it Collect-and-Collapse} and {\it Radiation-Driven Implosion}.

The {\it Collect-and-Collapse} mode \citep[hereafter C\&C; ][]{Elmegreen1977, Whitworth1994b, Dale2007, Dale2009, Wunsch2010} presupposes a rather homogeneous ambient medium; the expanding H{\sc ii} region then sweeps up this medium into a dense shell, which eventually becomes sufficiently massive to fragment gravitationally and form a new generation of stars \citep[e.g.][]{Wunsch2010}. One interesting feature of the C\&C mode is that it is expected to produce quite massive fragments, and therefore the possibility exists that these will spawn a new generation of massive stars, so that the process can repeat itself \citep[e.g.][]{Whitworth1994a}. In several regions (e.g. Sh2-212 \citep{Deharveng2008} and Sh-104 \citep{Deharveng2003}), there is observational evidence for massive star formation in shell-like structures very close to ionisation fronts. However, it has not yet been shown unequivocally that the formation of these massive stars has been triggered by C\&C.

\citet{Wunsch2012} analyse the dense clumps in the Carina flare and find evidence for shell fragmentation, which they explain with the {\bf P}ressure {\bf A}ssisted {\bf G}ravitational {\bf I}nstability \citep[PAGI; ][]{Wunsch2010} formalism, a model based on the C\&C scenario. With PAGI, the mass spectrum of the clumps forming in a swept-up, pressure-confined shell can be derived if the shell surface density and the confining pressure are known. However, \citet{Dale2011b} note that the clump mass spectrum is quickly changed due to oligarchic growth.

The {\it Radiation-Driven Implosion} mode \citep[hereafter RDI; ][]{Sandford1982, Bertoldi1989, Kessel2003} presupposes a rather structured ambient medium; the expanding H{\sc ii} region then advances most rapidly in those directions where the density is relatively low, thereby overtaking and compressing regions where the density is relatively high, and causing them to implode. Existing simulations of RDI have tended to focus on the onset and efficiency of triggered star formation in isolated pre-existing clumps \citep{Bisbas2011, Haworth2012}, and on the driving of turbulence \citep{Peters2011}. Three-dimensional simulations have been used to study the formation of bright rims and pillars \citep{Miao2006, Gritschneder2009, Gritschneder2010, Bisbas2011, Ercolano2011, Mackey2011} as well as the formation of ultra-compact \citep{MacLow2007,Peters2010, Peters2011}  and large-scale H{\sc ii} regions \citep{Mellema2006, Krumholz2007, Arthur2011, Walch2011}. 

\citet{Walch2011, Walch2012} have shown that in a realistic situation these two modes may operate in tandem: the expansion of the ionised gas acts {\it both} to organise the neutral gas into a range of dense structures, the most massive of which may be extended and shell-like, {\it and} to overrun these structures and compress them, so that they collapse, or collapse faster than they would otherwise have done. Consequently high- and low-mass fragments are formed and collapse, due to the {\bf E}nhancement of pre-existing {\bf D}ensity substructure and subsequent {\bf G}lobal {\bf I}mplosion' \citep[EDGI; ][]{Walch2011}. These conclusions are based on SPH simulations and observables derived by post-processing individual frames with radiative transfer. The initial structure of the cloud is assumed to be fractal. It is important to note that, although the ionising star may accelerate star formation in its vicinity (in the sense that star formation occurs sooner  than it would otherwise have done), and organises stars into distinct structures (partial shells, arcs and clusters), it actually reduces the net amount of star formation (as compared with the mass of stars that would eventually have formed in the absence of an ionising star). This is because the ionising star is very effective in dispersing a large fraction of the cloud gas (W12). Similar results have been found by \citet{Daleetal2012}.

Additional feedback processes like the protostellar jet and stellar wind of the central source, as well as radiation pressure, also add momentum and energy and could support the sweeping up of cold gas and the triggering of star formation. Protostellar jets are dynamically not important for triggering star formation as they are highly collimated, slower,  $v_{_{\mathrm JET}} \sim 300\;{\rm km/s}$, and overall less energetic than the radiative feedback of the massive star and its stellar wind. 
For the type of star considered here, the stellar wind is also likely to have a negligible effect. However, this estimate is more uncertain. Adopting the mass-luminosity scaling relation for main sequence stars, the central star has a mass of $\sim 25 \;{\rm M_\odot}$.  A star of this mass has a wind velocity of $\sim 3000\;{\rm km/s}$ and a mass loss rate of $\sim 10^{-7}\;{\rm M_\odot/yr}$ \citep{Ekstroem2012}. Therefore, the kinetic energy input of the wind is $\sim 9\times 10^{48}\;{\rm erg/Myr}$, which is low compared to the radiative energy input of $10^{52}$ erg/Myr. However, the momentum input of the wind may still be significant, as the conversion of radiative to kinetic energy is quite inefficient \citep{Walch2012}.
In paper I \citep{Walch2012}, we show that the cold shell material is accelerated to $\sim 7 {\rm km/s}$. Assuming a typical shell mass of $2000\; M_\odot$, the amount of radiative energy that has been converted to kinetic energy is $\sim 10^{48} {\rm erg/Myr}$. The wind bubble is usually confined within the HII region \citep{Weaver1977}, and thus the hot wind material shocks with the ionised gas that fills the bubble interior. It is unclear how much wind energy is radiated away during this process. In simulations of wind-bubble expansion caused by a $40\;{\rm M_\odot}$ star in uniform media, \citet{Toala2011} show that the fraction of the time-integrated mechanical wind energy, which is retained by the surrounding ISM in form of kinetic energy, is $\sim$ 10\%. Therefore the relative importance of the wind momentum input could be important. We will investigate this in a future paper.
Finally, radiation pressure on dust could add to the expansion of the cloud. In particular, radiation pressure might be the dominant feedback mechanism in the case of massive star clusters forming in giant molecular clouds with $> 10^6\;{\rm M}_\odot$ \citep{Murray2010}. However, \citet{KM2009} estimate that the impact of radiation pressure is small in HII regions driven by single or small N-clusters of massive stars. Therefore, it may be safely neglected in this study.

The plan of this paper is as follows. In Section \ref{SEC:METHOD} we describe  how we construct fractal molecular clouds and the numerical scheme that we use. In Section \ref{SEC:CLUMPS} we discuss the properties of the cold clumps swept up and/or compressed by an H{\sc ii} region expanding into a molecular cloud, as a function of its fractal dimension, ${\cal D}$. In Section \ref{SEC:STARS} we focus on the triggering of star formation, and characterise the distribution and the statistical properties of the stars formed, as a function of ${\cal D}$. We summarise our main conclusions in Section \ref{SEC:CONC}. 

\section{Numerical method \& Initial conditions}\label{SEC:METHOD}%

\subsection{Generation of fractal molecular clouds}\label{fractal}%

It appears that the internal structure of molecular clouds is broadly self-similar over four orders of magnitude, from $\sim 0.1\,{\rm pc}$ to $500\,{\rm pc}$ \citep[e.g.][]{Bergin2007, Sanchez2010}, and that it is approximately fractal, with dimension in the range $2.0\!\la\!{\cal D}\!\la\!2.8$ \citep[e.g.][]{Falgarone1991, Elmegreen1996, Stutzki1998, Vogelaar1994, Lee2004, Sanchez2005, Sanchez2007, Schneider2011, Miville2010}.

To construct a fractal cloud we consider a $\,2\!\times\!2\!\times\!2\,$ cubic computational domain, and specify three parameters: the fractal dimension, ${\cal D}\!=\!2.0,\,2.2,\,2.4,\,2.6,\,2.8$ (or equivalently \citep{Stutzki1998} the spectral index, $n=8-2{\cal D}=4.0,\,3.6,\,3.2,\,2.8,\,2.4$, where $n$ relates to the 3D density power spectrum, $P_k\propto k^{-n}$, $k$ is wavenumber, and $k\!=\!1$ corresponds to the linear size of the cubic domain, i.e. $\lambda(k)=2/k$); a random seed, ${\cal R}$, which allows us to generate multiple realisations; and a density-scaling parameter, $\rho_{_{\rm O}}$. We populate all modes having integer $k_x$, $k_y$ and $k_z$ in $(1,128)$, with random phases and amplitudes drawn from the power-spectrum. Next we perform an FFT to evaluate the function $\rho_{_{\rm FFT}}(x,y,z)$ on a $128^3$ Cartesian grid spanning the computational domain. Finally we compute the density in the computational domain, according to
\begin{equation}
\rho(x,y,z)=\exp{\left(\frac{\rho_{_{\rm FFT}}(x,y,z)}{\rho_0} \right)}\,,\label{EqScale}
\end{equation}
where $\rho_{_{\rm O}}$ is a dimensionless scaling parameter \citep{Shadmehri2011}. SPH particles are then distributed randomly in each cell of the grid according to its density, and SPH particles that fall outside a unit-radius sphere are culled. The resulting sphere can then be re-scaled to arbitrary total mass, $M_{_{\rm C}}$ and arbitrary radius, $R_{_{\rm C}}$.

With the above procedure, the random seed, ${\cal R}$ completely determines the pattern of the density field. The fractal dimension determines the distribution of power: for small ${\cal D}$ (large $n$), most of the power is on large scales, so the density field is dominated by extended structures; conversely, for large ${\cal D}$, (small $n$), there is more power on small scales, and so the density field is dominated by smaller structures. The scaling parameter, $\rho_{_{\rm O}}$, determines the density contrast, and hence the width of the (approximately log-normal) density PDF; increasing $\rho_{_{\rm O}}$ decreases the density contrast and therefore reduces the width of the PDF.

\subsection{Numerical method}
We use the SPH code \textsc{seren} \citep{Hubber2011}, which is well-tested and has already been applied to many problems in star formation \citep[e.g.][]{Walch2011, Bisbas2011, Stamatellos2011}. We employ the SPH algorithm of \citet{Monaghan1992} with a fixed number of neighbours, $N_{_{\rm NEIGH}}=50$. The SPH equations of motion are solved with a second-order Leapfrog integrator, in conjunction with an hierarchical block time-stepping scheme. Gravitational forces are calculated using an octal spatial decomposition tree \citep{Barnes1986}, with monopole and quadrupole terms and a Gadget-style opening-angle criterion \citep{Springel2001}. We use the standard artificial viscosity prescription \citep{Monaghan1983}, moderated with a Balsara switch \citep{Balsara1995}.

The ionizing radiation is treated with an HEALPix-based, adaptive ray-splitting algorithm, which allows for optimal resolution of the ionization front in high resolution simulations \citep[see ][]{Bisbas2009}. Along each HEALPix ray, the radiative transfer is evaluated at discrete points, $j$. These points are separated by $f_1h_j$, where $f_1=0.6$ is an accuracy parameter and $h_j$ is the local smoothing length (adjusted to enclose $\sim$50 SPH neighbours). A ray is split if the linear separation of neighbouring rays is greater than $f_2h_j$, where $f_2$ is the angular resolution parameter. Typically, good results are achieved for $1.0\leq f_2\leq 1.3$; here we use $f_2=0.8$ to further increase the angular resolution of the ray tracing scheme. We allow for a maximum number of $l_{_{\rm MAX}}=11$ HEALPix levels corresponding to $12 \times 4^{l_{_{\rm MAX}}} \approx 5 \times 10^7$ rays if the whole sphere were refined to $l_{_{\rm MAX}}$. In the simulations presented here, most directions only require $\leq 8$ levels of refinement, and there are typically $\sim 10^5$ rays in total.

The gas is assumed to be either fully molecular with a mean molecular weight of $\mu_{_{\rm NEUT}}=2.38$, or fully ionized with $\mu_{_{\rm ION}}=0.7$. The temperature of ionized gas is set to $T_{_{\rm ION}}=$10,000 K. The temperature of neutral gas is given by a barotropic equation of state,
\begin{equation}
  T(\rho)=T_{_{\rm NEUT}} \left[ 1+(\rho/\rho_{_{\rm CRIT}})^{(\gamma-1)}\right]\,,
\end{equation}
where $T_{_{\rm NEUT}}=30\,\rm{K}$, $\rho_{_{\rm CRIT}}=10^{-13}\,{\rm g}\,{\rm cm}^{-3}$, and $\gamma=5/3$. The use of $T_{_{\rm NEUT}}=30\,{\rm K}$ may influence the fragmentation properties of the forming shell. Since the shell becomes very dense and should therefore be allowed to cool further, it might fragment more efficiently than currently seen in our simulations; for this reason we will explore a more complicated cooling function in a future paper.

We introduce sinks at density peaks above $\rho_{_{\rm SINK}}=10^{-11}\,\rm{g}\,\rm{cm}^{-3}$, using the new algorithm developed by \citet{Hubber2013}. Since $\rho_{_{\rm SINK}}\!\gg\!\rho_{_{\rm CRIT}}$, a condensation that is transformed into a sink is normally already well into its Kelvin-Helmholtz contraction phase. Once formed, a sink is able to accrete gas smoothly from its surroundings and thereby grow in mass.

\begin{table}
\begin{center}
\begin{tabular}{ccccccc}\hline
ID & ${\cal D}$ & Seed & $\bar{\rho}_{_{\rm MW}}$ & $\sigma_{_{\rm O}}^2$ & $t_{_1}$ & $t_{_{15}}$ \\
 & & & $\overline{10^{-21}{\rm g\,cm}^{-3}}$ & & $\overline{\rm Myr}$ & $\overline{\rm Myr}$ \\\hline
${\cal D}$2.0/O7(1) & 2.0 & 1 & 1.23 & 1.08 & 0.46 & 0.66 \\
${\cal D}$2.0/O7(2) & 2.0 & 2 & 1.17 & 1.42 & 0.50 & 0.61 \\
${\cal D}$2.0/O7(3) & 2.0 & 3 & 1.17 & 1.10 & 0.56 & 0.66 \\\\
${\cal D}$2.2/O7(1) & 2.2 & 1 & 1.17 & 1.06 & 0.47 & 0.62 \\
${\cal D}$2.2/O7(2) & 2.2 & 2 & 0.98 & 1.12 & 0.43 & 0.61 \\
${\cal D}$2.2/O7(3) & 2.2 & 3 & 1.17 & 0.90 & 0.57 & 0.66 \\\\
${\cal D}$2.4/O7(1) & 2.4 & 1 & 1.07 & 0.76 & 0.51 & 0.67 \\
${\cal D}$2.4/O7(2) & 2.4 & 2 & 0.93 & 0.79 & 0.35 & 0.66 \\
${\cal D}$2.4/O7(3) & 2.4 & 3 & 0.93 & 0.55 & 0.47 & 0.87 \\\\
${\cal D}$2.6/O7(1) & 2.6 & 1 & 0.89 & 0.53 & 0.45 & 1.00 \\
${\cal D}$2.6/O7(2) & 2.6 & 2 & 0.89 & 0.58 & 0.44 & 0.72 \\
${\cal D}$2.6/O7(3) & 2.6 & 3 & 0.85 & 0.44 & 0.45 & 0.81 \\\\
${\cal D}$2.8/O7(1) & 2.8 & 1 & 0.81 & 0.41 & 0.67 & 0.97 \\
${\cal D}$2.8/O7(2) & 2.8 & 2 & 0.81 & 0.44 & 0.46 & 0.73 \\
${\cal D}$2.8/O7(3) & 2.8 & 3 & 0.85 & 0.36 & 0.71 & 0.86 \\\hline
\end{tabular}
\caption{Simulation parameters. Column 1 gives the simulation ID; column 2, the fractal dimension (${\cal D}$); column 3, the random seed used; column 4, the mass-weighted mean-density ($\bar{\rho}_{_{\rm MW}}$); column 5, the variance of the log-normal mass-weighted density PDF ($\sigma_{_{\rm O}}^2$); column 6 the time at which the first sink ("protostar") forms ($t_1$); and column 7, the time at which the fifteenth sink forms ($t_{15}$).}
\label{TAB:PARAMS}
\end{center}
\end{table}

\subsection{Initial conditions}

We consider a spherical cloud with total mass ${\rm M}_{_{\rm MC}}=10^4\,{\rm M}_\odot$, radius $R_{_{\rm MC}}=6.4\,\rm{pc}$, mean density $\bar{\rho}=6.17 \times 10^{-22}\,{\rm g}\,{\rm cm}^{-3}$, and mean freefall time $t_{_{\rm FF}}=3\,{\rm Myr}$, with an O7 star at its centre emitting Lyman continuum photons at a rate $\dot{\cal N}_{_{\rm LyC}}=10^{49}\,{\rm s}^{-1}$. We treat five fractal dimensions, ${\cal D}=2.0,\,2.2,\,2.4,\,2.6\,{\rm and}\,2.8$, corresponding to spectral indices $n=4.0,\,3.6,\,3.2,\,2.8\,{\rm and}\,2.4$ respectively. For lower ${\cal D}$ (higher $n$), a larger fraction of the power is invested in the extended structures. For each value of ${\cal D}$, we treat three different realisations, by invoking three different random seeds, ${\cal R}$, but we use the same three random seeds. Simulations with the same seed, ${\cal R}$, but different fractal dimension, ${\cal D}$, have the same pattern of density peaks and troughs, and differ only in the sense that for lower ${\cal D}$ the more extended structures exhibit more density contrast, and the more compact structures exhibit less density contrast. We use a single value of the scaling parameter, $\rho_{_{\rm O}}=1.0$. By using three different random seeds, we are able both to evaluate the extent to which the results depend on the particular choice of ${\cal R}$, and to improve the statistics of the results for each individual value of ${\cal D}$. The simulations are given IDs of the form "${\cal D}$2.0/O7(1)", where characters two through four (following ${\cal D}$) give the fractal dimension, characters six and seven (following the oblique stroke) indicate that the cloud is ionised by an O7 star, and character nine (in parentheses) records which seed was used. All simulations are preformed with ${\cal N}_{_{\rm TOT}}\!\sim\!2.5\times 10^6$ particles, and therefore the minimum mass that can be resolved is $M_{_{\rm MIN}}\!=\!50\,M_{_{\rm MC}}/{\cal N}_{_{\rm TOT}}\!\sim\!0.2\,{\rm M}_{_\odot}$.

\begin{figure}
\includegraphics[width=92mm, angle=0]{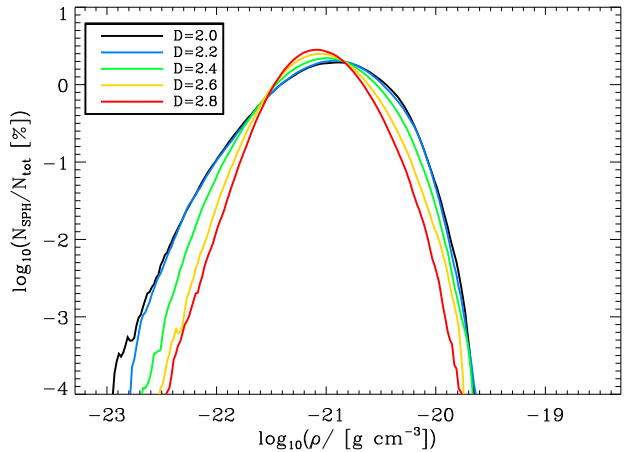} 
\caption{The mass-weighted logarithmic density PDFs for the initial conditions generated with different fractal dimensions (${\cal D}=2.0,\,2.2,\,2.4,\,2.6,\,2.8$) but the same random seed (seed 1). The mass-weighted mean densities and logarithmic variances for these distributions (and those obtained with the other two seeds) are given in Table \ref{TAB:PARAMS}.} 
\label{FIG:RHOPDF}
\end{figure}

\begin{figure*}
\begin{center}
\begin{tabular}{c}
\includegraphics[width=162mm]{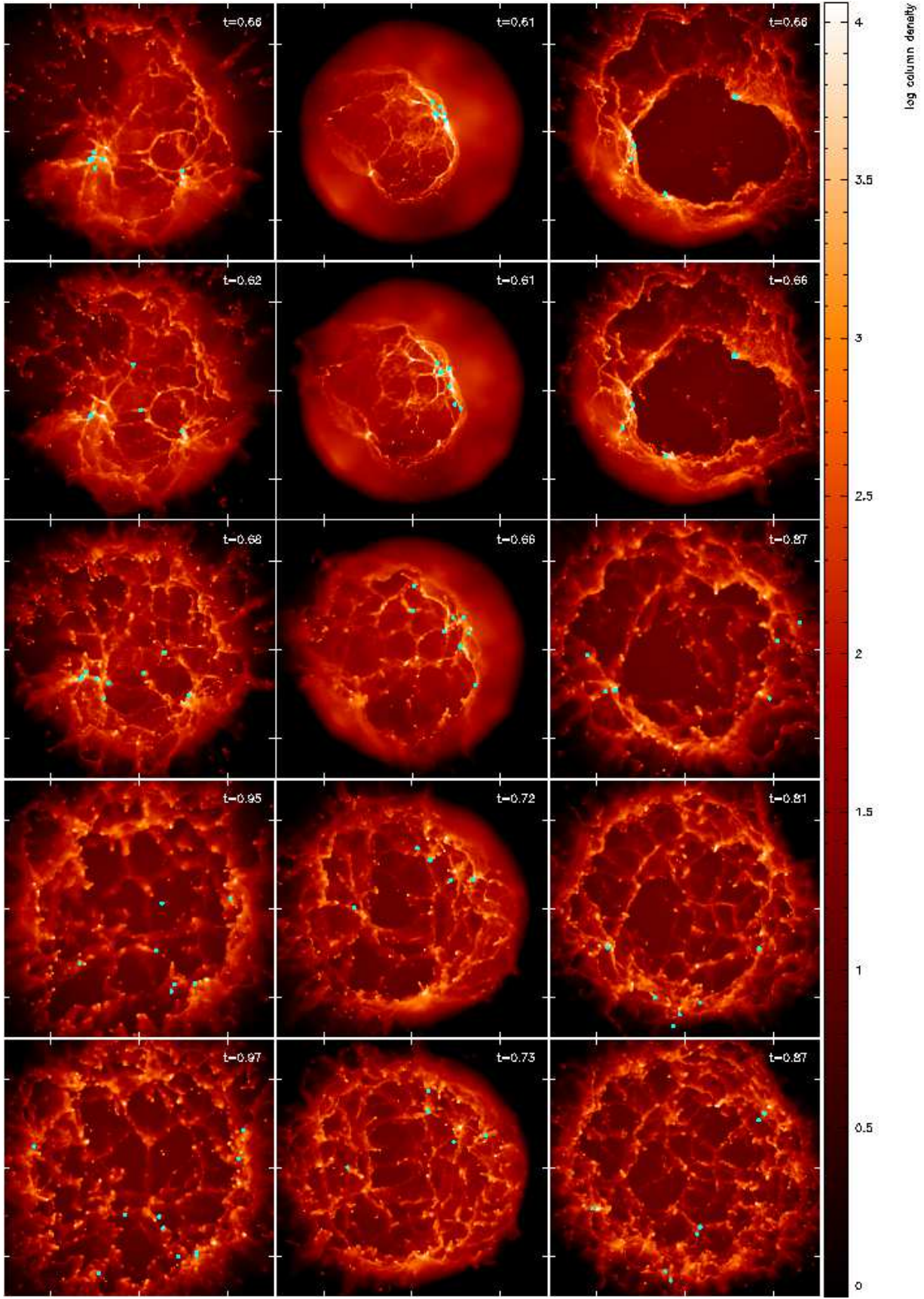}
\end{tabular}
\caption{False-colour column density images of all simulations at $t_{15}$ (see Table \ref{TAB:PARAMS}). The ionising source is located at the center of each panel of size $14 \times 14$ pc. From top to bottom ${\cal D}$ increases from ${\cal D}=2.0$ (top) to ${\cal D}=2.8$ (bottom). Each column represents clouds generated with the same random seed, and hence an initial density field with the same pattern. Sink particles are marked as turquoise dots; many of the sinks are in close (unresolved) multiple systems.}
\label{FIG_M1}
\end{center}
\end{figure*}

Fig. \ref{FIG:RHOPDF} shows the mass-weighted density PDFs for the clouds created with the first seed. Since they are approximately log-normal, we can compute a standard deviation, $\sigma_{_{\rm O}}$, for each one. Evidently $\sigma_{_{\rm O}}$ increases with decreasing ${\cal D}$, because at lower ${\cal D}$ a larger fraction of the power is concentrated in a few large-scale structures. This in turn means that for lower ${\cal D}$ there is, at the outset, more gas at large densities, and therefore star formation tends to occur sooner. Table \ref{TAB:PARAMS} gives the basic parameters for the complete suite of simulations.

\section{Spatial distribution and intrinsic statistics of clumps}\label{SEC:CLUMPS}%

\subsection{General morphology}%

The morphology of the evolving H{\sc ii} region is strongly dependent on the fractal dimension, ${\cal D}$, of the initial molecular cloud. Fig. \ref{FIG_M1} shows false-colour column-density images of all 15 simulations (i.e. from top to bottom, all five values of ${\cal D}$, and from left to right, all three random seeds), projected onto the $z\!=\!0$ plane. The snapshots are all taken at time $t_{15}\,$, i.e. the time at which the fifteenth sink is created, and values of $t_{15}$ are given in Table \ref{TAB:PARAMS}. Sinks are shown as turquoise dots. Due to the variance in the initial density fields, different realisations with the same fractal dimension exhibit different star formation rates, but there is a tendency for star formation to occur later when ${\cal D}$ is higher. \citet{Walch2012} report a systematic morphological transition, from a shell-dominated H{\sc ii} region structure for low fractal dimension (${\cal D}=2.0\,{\rm or}\,2.2$), to a pillar-dominated structure at high fractal dimension (${\cal D}=2.6\,{\rm or}\,2.8$). Here, we confirm that this morphological transition persists when using different realisations, even though the detailed appearance of an H{\sc ii} region depends strongly on the random seed used, and on the viewing angle. In the following, we utilise the improved statistics afforded by multiple realisations to investigate the characteristics of the cold, swept-up clumps bordering the H{\sc ii} region and the subsequent triggered star formation, as a function of ${\cal D}$.

\begin{figure*}
\begin{center}
\begin{tabular}{c}
\includegraphics[width=180mm]{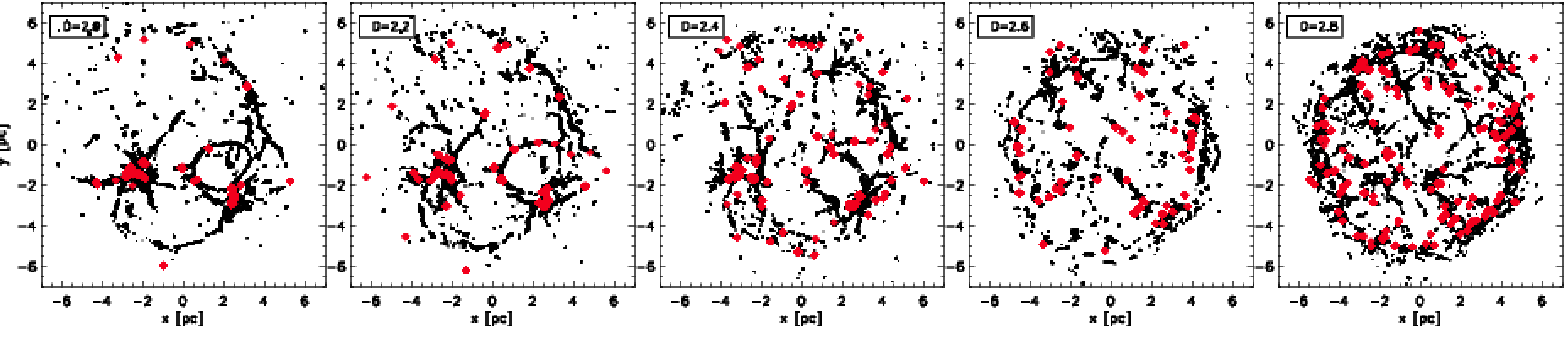}
\end{tabular}
\caption{The positions of high-density SPH particles, projected onto the $z\!=\!0$ plane, for all the simulations set up with the first seed (i.e. ${\cal D}2.0/O7(1)$ (left) to ${\cal D}2.8/O7(1)$ (right) at $t_{_{\rm 15}}$. Particles, $p$, with density $6\times 10^{-20}\,{\rm g\,cm}^{-3}<\rho_p<6\times 10^{-19}\,{\rm g\,cm}^{-3}$ are plotted in black, and those with density $\rho_p>6\times 10^{-19}\,{\rm g\,cm}^{-3}$ are plotted in red. The red particles are the 'core' particles used to determine the core statistics.}
\label{FIG:PARTDIST}
\end{center}
\end{figure*}

\begin{figure*}
\begin{tabular}{ll}
\includegraphics[width=92mm]{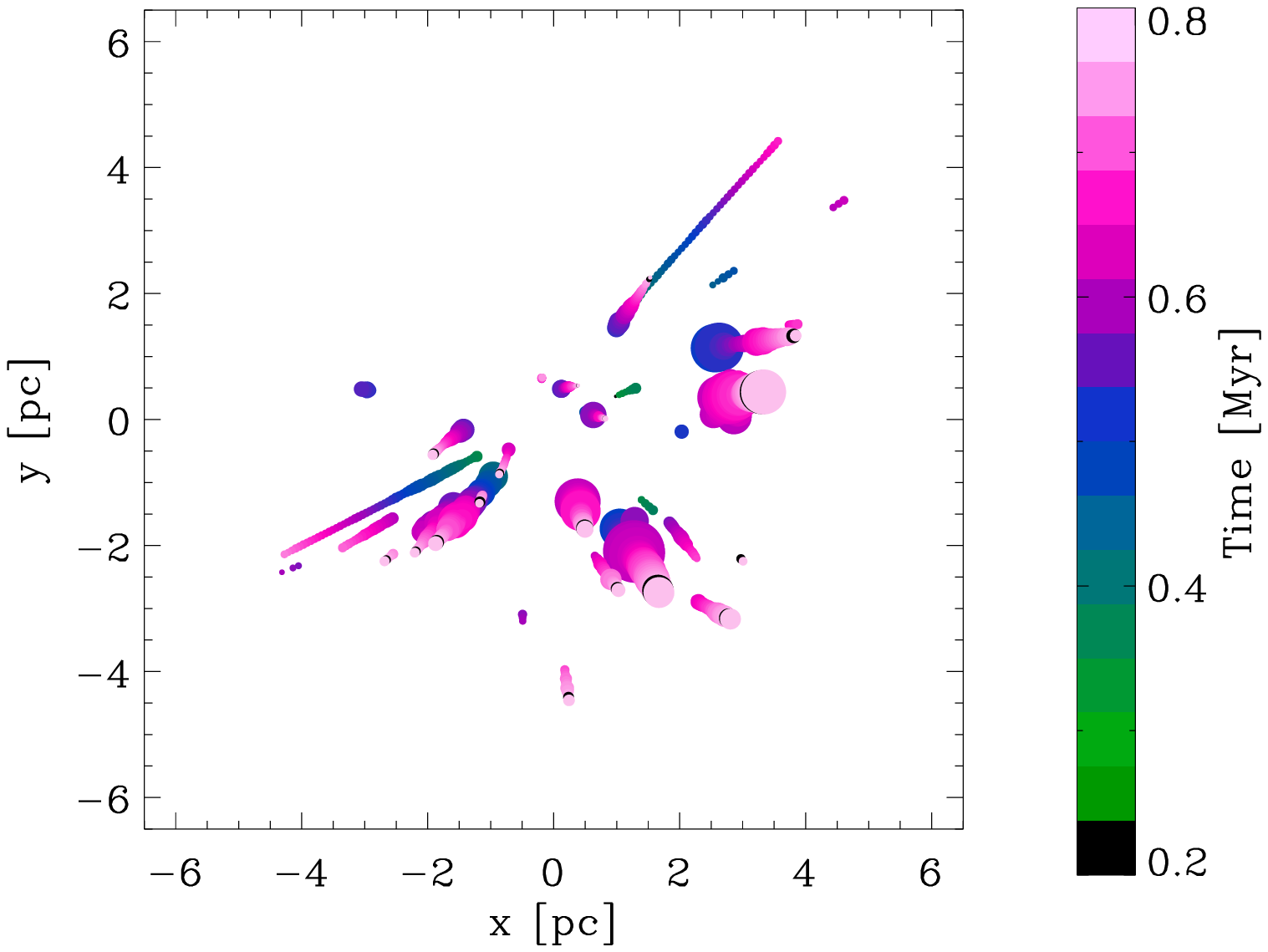} &
\includegraphics[width=92mm]{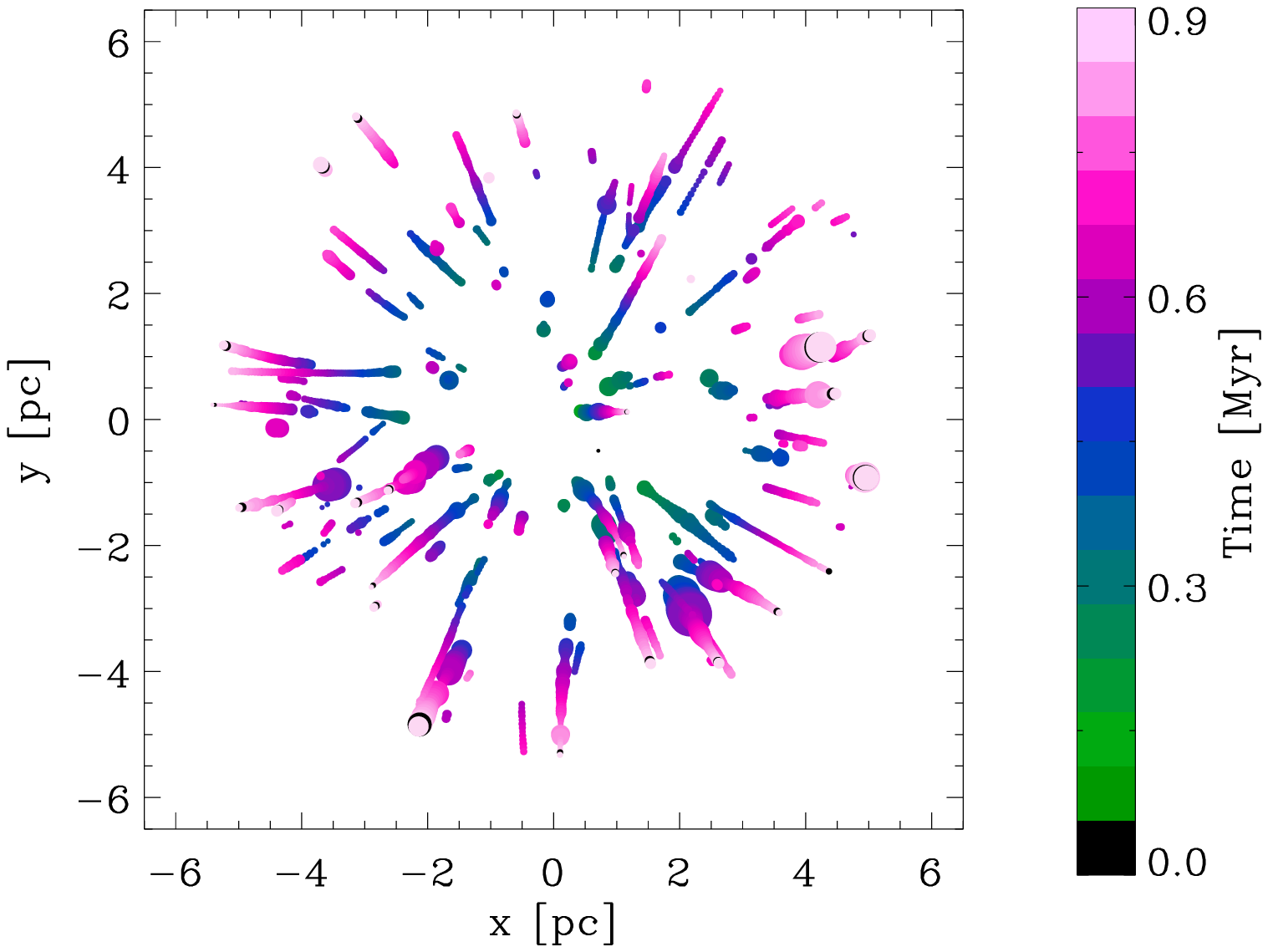} \\
\end{tabular}
\caption{The dynamical evolution of clump masses and positions projected on the $z\!=\!0$ plane. False colour encodes time (see colour bar), and track width encodes clump mass. The left frame shows a low fractal dimension case (${\cal D}$2.0/O7(1)), and the right frame a high fractal dimension case (${\cal D}$2.8/O7(1)).}
\label{FIG:CLUMPS}
\end{figure*}

\subsection{Clump formation in shells and pillars}%

There is strong observational evidence that H{\sc ii} regions are usually surrounded by shell-like structures that contain dense, molecular clumps, and that these clumps are often the sites of new star formation \citep[e.g.][]{Zavagno2010, Deharveng2010}. However, it is still unclear how these clumps form, i.e. what is the relative importance of RDI, C\&C, PAGI, EDGI, etc. (see Section 1). In some regions the clump mass function (CMF) appears to be rather similar to the one found in low-mass star forming regions \citep{Wunsch2012}, whereas in other regions the presence of more massive clumps suggest that the C\&C mechanism may have been at work \citep{Deharveng2003}. 

To investigate the statistics of the cold clumps formed in our simulations, we first identify all the SPH particles having density $\rho_p>6\times 10^{-19}\,{\rm g\,cm}^{-3}$ (i.e., for molecular gas, $n_{{\rm H}_2}>1.5\times 10^5\,{\rm cm}^{-3}$). This density is sufficiently high for the gas to couple thermally to the dust, and a significant proportion of it should be destined to form stars. The free-fall time for a lump of gas with a uniform density of $6\times 10^{-19}\,{\rm g\,cm}^{-3}$ is $\sim$0.1 Myr. The positions of the SPH particles selected in this way are plotted in Fig. \ref{FIG:PARTDIST} for all the simulations performed with the first random seed; SPH particles having density $\rho_p>6\times 10^{-19}\,{\rm g\,cm}^{-3}$ are plotted in red, and -- for comparison -- those having density $6\times 10^{-20}\,{\rm g\,cm}^{-3}<\rho_p<6\times 10^{-19}\,{\rm g\,cm}^{-3}$ in black. Individual clumps are clearly picked up more reliably with the higher density threshold.

We identify individual clumps by applying the Friends-of-Friends (hereafter FoF) algorithm to the selected subset of high-density SPH particles, using a linking length of $\ell=0.05\,{\rm pc}$. Thus a clump represents a collection of high-density SPH particles, all of which are no further than $0.05\,{\rm pc}$ from at least one other member of this collection; any value of $\ell$ in the range $(0.01,0.1){\rm pc}$ delivers broadly similar clumps statistics.

Fig. \ref{FIG:CLUMPS} illustrates the dynamical evolution of the clumps in the simulations ${\cal D}$2.0/O7(1) and ${\cal D}$2.8/O7(1); positions are projected on the the $z\!=\!0$ plane, time is colour-coded, and the width of the symbol encodes the mass of the clump. In the case with low fractal dimension (left-hand frame, ${\cal D}\!=\!2.0$), the plot is dominated by a small number of massive clumps, which are distributed very anisotropically with respect to the ionising star. In the case with high fractal dimension (right-hand frame, ${\cal D}\!=\!2.8$), there are many more clumps, but they are much less massive, and they are distributed much more isotropically. All the clumps are being driven outwards by the rocket effect \citep{Kahn1954, Oort1955}, at speeds up to $\sim\!10\,{\rm km\,s}^{-1}$.

\begin{figure}
\includegraphics[width=92mm]{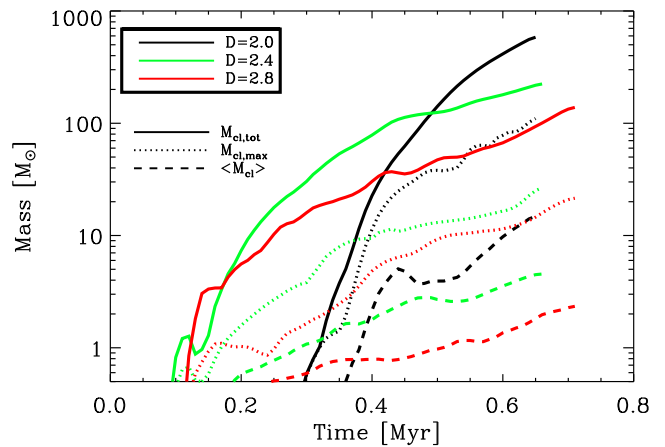}
\caption{Time evolution of the total mass in clumps (solid line), the maximum clump mass (dotted line), and the mean clump mass (dashed line), for ${\cal D}=2.0$ (black), ${\cal D}=2.4$ (green) and ${\cal D}=2.8$ (red). In each case, the results from three different realisations have been collated. The results for ${\cal D}=2.2$ and ${\cal D}=2.6$ are not included, simply to make the plot easier to read.}
\label{FIG:MCLUSTER}
\end{figure}

Fig. \ref{FIG:MCLUSTER} illustrates the time evolution of clump masses. Different ${\cal D}$ are represented by different colours. For a given ${\cal D}$, the solid line shows the total mass in clumps; the dotted line shows the mass of the most massive clump; and the dashed line shows the mean clump mass. As ${\cal D}$ is increased, clumps start forming earlier, but ultimately the total mass in clumps and the masses of individual clumps are lower. To make the plot easier to read we only present results for ${\cal D}=2.0,\,2.4\,{\rm and}\,2.8$, but the in-between values show the same trends; for each of these ${\cal D}$-values, we have combined the results from all three realisations.

The growth of a clump is not driven by gravity \citep[as, for example, in Bondi accretion;][]{Bondi1952}, and there is also no strictly oligarchic growth \citep[as e.g. inferred by][]{Dale2011b}. Rather, in the first instance, matter is driven into a clump wherever the shock waves that precede the expanding ionisation front converge. Later on the ionisation front will start to erode the clump, but at the same time the clump will be driven outwards by the rocket effect, and sweep up material from further out in the cloud, like a snowplough. The evolution of its mass is then a competition between these two effects.

\subsection{Clump statistics}%

\begin{figure*}
\begin{center}
\begin{tabular}{cc}
\includegraphics[width=85mm]{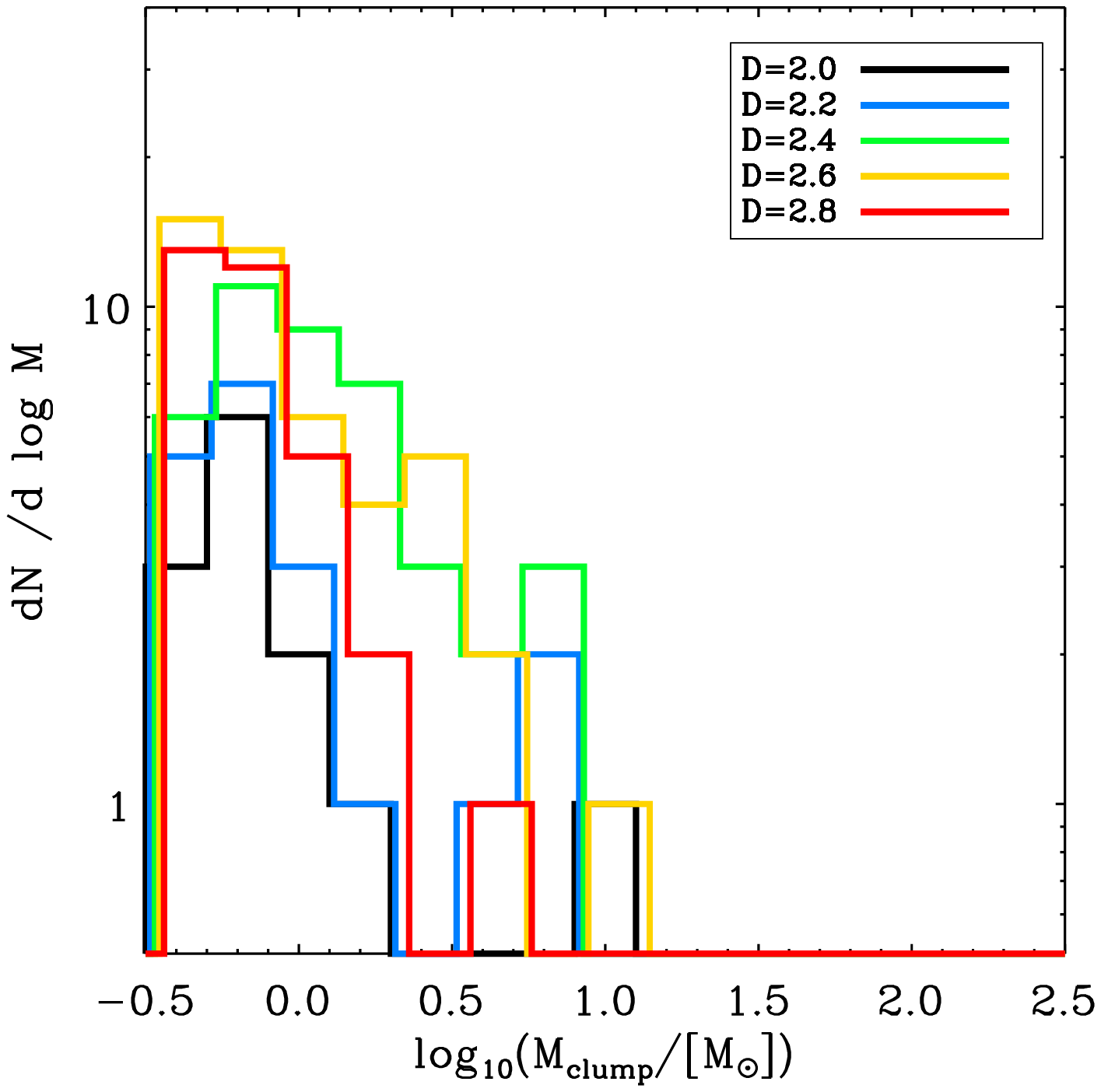} &
\includegraphics[width=85mm]{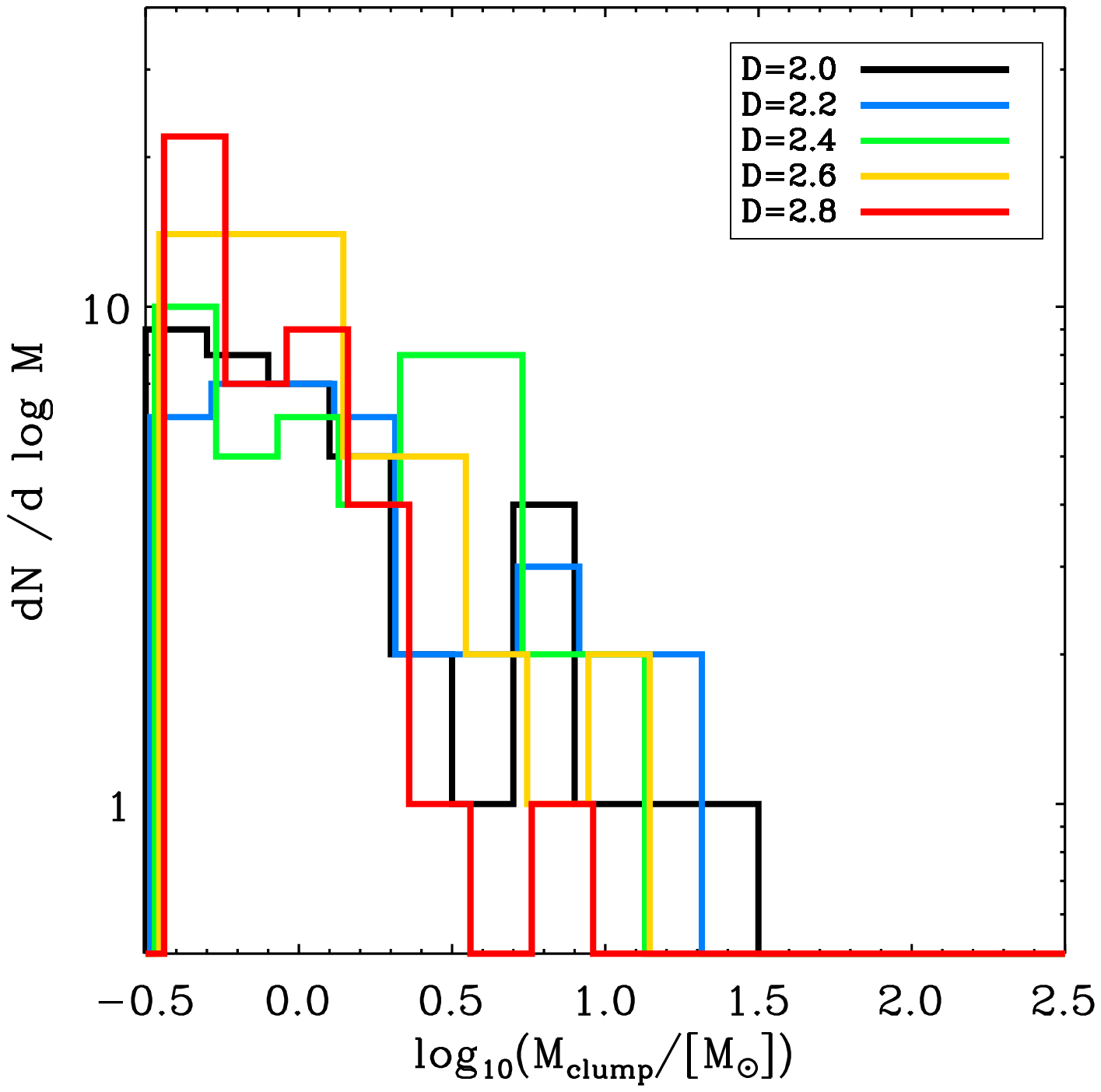} \\
\includegraphics[width=85mm]{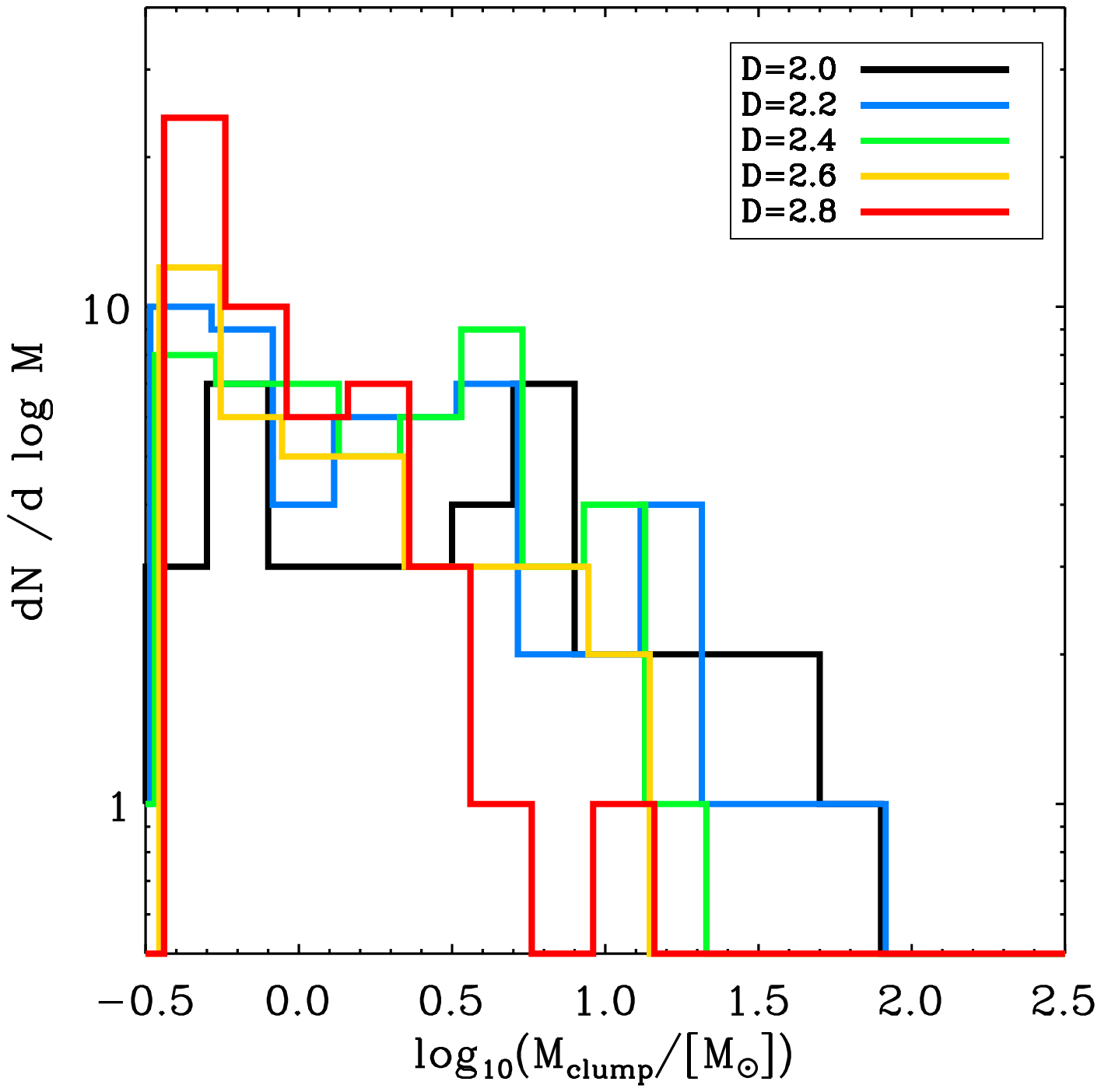} &
\includegraphics[width=85mm]{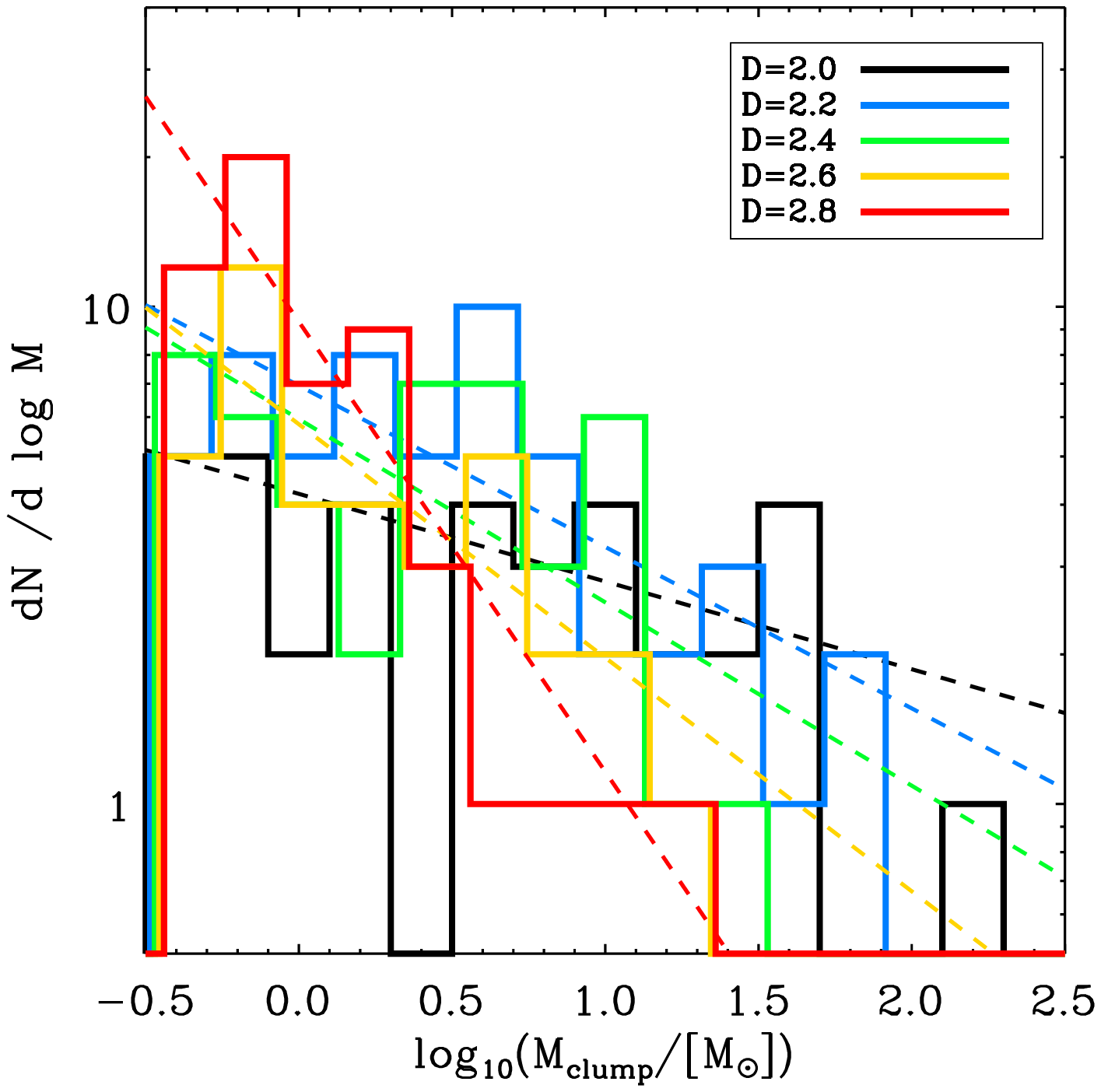}
\end{tabular}
\caption{CMFs at times, $t=0.4,\,0.5,\,0.6\,{\rm and}\,0.66$ Myr (reading from top left to bottom right). For each ${\cal D}$, we have combined the clump masses from the three different realisations. At $t=0.66$ Myr we determine the power law slopes of the CMFs by means of $\chi^2$-minimisation (dashed lines).} 
\label{FIG:CMF}
\end{center}
\end{figure*}

Fig. \ref{FIG:CMF} displays clump mass functions (CMFs) for the different ${\cal D}$ values, at times $t=0.40,\,0.50,\,0.60\,{\rm and}\,0.66\,{\rm Myr}$; $\;0.66\,{\rm Myr}$ is the last time reached by all simulations. The bin size is $\Delta \log_{10}\left(M/{\rm M}_{_\odot}\right)\!=\!0.2$, and the plot is logarithmic (so that the Salpeter stellar IMF would have slope $m=-1.35$). Typically, the mass function of cold clumps -- which are not necessarily gravitationally bound -- is somewhat flatter than Salpeter, with slopes $m\sim -0.7$ being commonly reported \citep[e.g.][]{Kramer1998, Wunsch2012}. However, a steeply decreasing power law is not recovered in all molecular clouds; for example, in Orion, \citet{Li2007} even report an {\it increasing} power law of $m=+0.15$ over the mass range $0.1 \;{\rm M}_\odot \le M \le 10\;{\rm M_\odot}$.

At early times all CMFs appear quite similar, but as time advances, there are three main trends. First, more clumps are formed, as more gas is swept up by the expanding H{\sc ii} region. Second, the CMFs extend to higher mass, as a few of them have trajectories that cause them to intercept and sweep up dense material as they plough outwards. Third, the CMFs are shallow, almost flat, for ${\cal D}=2.0$, and become increasingly steep with increasing ${\cal D}$. We fit the CMFs at $t=0.66\,{\rm Myr}$ with a power-law, using $\chi^2$-minimisation; the resulting fits are shown as dashed lines on Fig. \ref{FIG:CMF}, and have the following slopes:
\begin{center}
\begin{tabular}{rccccc}
${\cal D}\;=\;$ & 2.0   & 2.2   & 2.4   & 2.6   & 2.8   \\
$m\;=\;$        & -\,0.18 & -\,0.32 & -\,0.37 & -\,0.47 & -\,0.91 \\
\end{tabular}
\end{center}
The principal cause of this difference is that low ${\cal D}$ delivers coherent, extended density enhancements in the initial cloud, and this promotes the growth of high-mass clumps within the shells bordering the expanding H{\sc ii} region; although these large clumps are eroded by the ionisation front on the side facing the star, they are also pushed outwards sweeping up the large amounts of neutral gas on the side facing away from the ionising star. In contrast, high ${\cal D}$ delivers smaller structures in the initial density field, and these develop into cometary globules and pillars; the ionisation front wraps round them, and so they are eroded from many different directions and having small cross-sections they do not sweep up so much extra mass as they plough outwards.

\subsection{Internal velocity dispersion}%

\begin{figure}
\includegraphics[width=90mm]{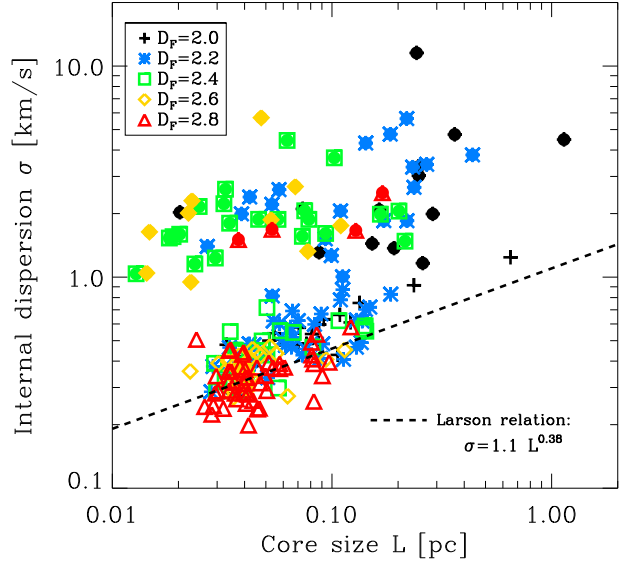} \\
\includegraphics[width=90mm]{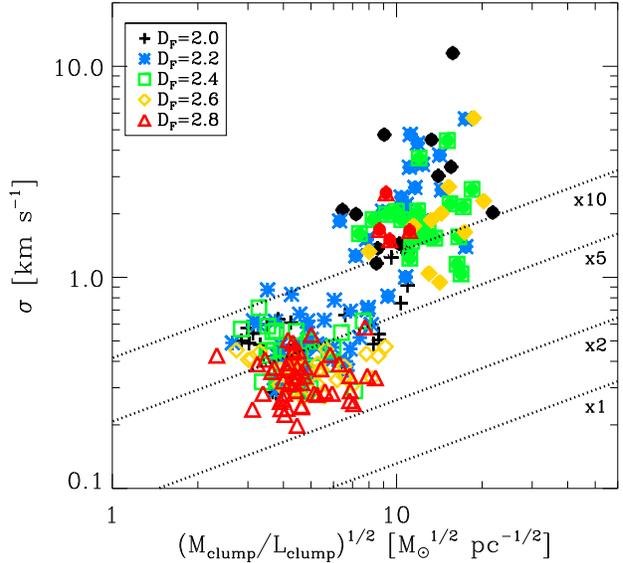} 
\caption{{\sc Top:} The internal velocity dispersion $\sigma$ of a clump plotted against its maximum extent $L$, at $t=0.66\,{\rm Myr}$. Star-forming clumps are marked with an additional filled circle in the same colour. We also show Larson's line-width-size relation (dashed line). {\sc Bottom:} Velocity dispersion $\sigma$ as a function of the square-root of the clump mass-to-size ratio, $(M_{_{\rm CLUMP}}/L_{_{\rm CLUMP}})^{1/2}$. The relationship derived for galactic molecular clouds is indicated by the lowest dotted line \citep[][ see Eq. \ref{EQ:HEYER}]{Heyer2009}. The following dotted lines indicate the \citet{Heyer2009} relation multiplied by a factor of 2, 5, and 10. }
\label{FIG:DISPSIZE}
\end{figure}

On Fig. \ref{FIG:DISPSIZE} (top panel) we plot the velocity dispersion inside each clump, $\sigma_{_{\rm CLUMP}}$, against its linear size, $L_{_{\rm CLUMP}}$, at time $t\!=\!0.66\,{\rm Myr}$. The velocity dispersion of a clump is computed by adding in quadrature the contribution from non-thermal motions (i.e. the velocity dispersion of the constituent SPH particles) and the contribution from thermal velocity dispersion. The linear size of a clump is simply its maximum extent. Many of the clumps conform approximately to Larson's Scaling relation \citep{Larson1981},
\begin{equation}\label{EQ:LARSON}
\frac{\sigma_{_{\rm CLUMP}}}{\rm km/s} = 1.1\; \left (\frac{L_{_{\rm CLUMP}}}{\rm pc}\right)^{\;0.38}\,.
\end{equation}
However, the clumps that are forming stars (hereafter "star-forming clumps", represented by symbols containing a filled circle) all lie well above Larson's Scaling Relation. Their velocity dispersion is higher because they are being accelerated and compressed by the H{\sc ii} region, and because they contain high-velocity flows onto forming protostars. 

\citet{Heyer2009} have suggested that the velocity dispersions in molecular clouds might be fitted more accurately if one assumes approximate virialisation, i.e. $\sigma_{_{\rm CLUMP}}\simeq(\pi G\Sigma_{_{\rm CLUMP}} R_{_{\rm CLUMP}}/5)^{1/2}$, where $\Sigma_{_{\rm CLUMP}}$ and $R_{_{\rm CLUMP}}$ are, respectively, the surface-density and radius of a clump (the triggered clumps typically have $10^2 \stackrel{<}{\sim}\Sigma_{_{\rm CLUMP}} \stackrel{<}{\sim}10^4\;{\rm M_\odot/pc^2}$). Putting $\Sigma_{_{\rm CLUMP}}\!=\!M_{_{\rm CLUMP}}/\pi R^2_{_{\rm CLUMP}}$ and $R_{_{\rm CLUMP}}\!=\!L_{_{\rm CLUMP}}/2$, this reduces to
\begin{equation}\label{EQ:HEYER}
\sigma_{_{\rm CLUMP}}\simeq \left(\frac{2GM_{_{\rm CLUMP}}}{5L_{_{\rm CLUMP}}}\right)^{1/2}\,.
\end{equation}
\vspace{-0.1cm}
In the lower panel of Fig. \ref{FIG:DISPSIZE} we plot $\sigma_{_{\rm CLUMP}}$ against $(M_{_{\rm CLUMP}}/L_{_{\rm CLUMP}})^{1/2}$. We find that all clumps lie above the Heyer Scaling Relation (Eqn. \ref{EQ:HEYER}), but those that are forming stars have higher $\sigma$-values, more than a factor $\sim\! 10$ higher.
\begin{figure}
\begin{tabular}{l}
\includegraphics[width=92mm]{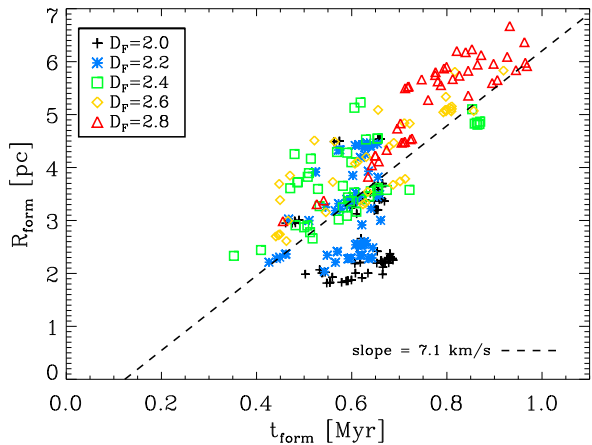} \\
\end{tabular}
\caption{The radial distance from the ionising source at sink formation, $R_{_{\rm FORM}}$, as a function of the sink formation time, $t_{_{\rm FORM}}$ for all simulations. Results obtained with different ${\cal D}$ are represented with different colours and different symbols. The best linear fit to this distribution of sinks is obtained using a slope of $7.1\;{\rm km/s}$ (dashed line).}
\label{FIG:MSINK}
\end{figure}

\section{Spatial distribution and intrinsic statistics of stars}\label{SEC:STARS}%

All the star formation in the simulations presented here is triggered. This has been demonstrated unequivocally in \citet{Walch2012} where, for comparison, we evolve the same fractal clouds without a central ionising star, and show that spontaneous star formation does not occur until long after the evolution time considered here. To characterise the consequences of triggered star formation, we collate the statistical properties of the sink particles, as a function of ${\cal D}$. We employ the improved sink particle algorithm of \citet{Hubber2013}, which has been demonstrated to improve the robustness of sink particle properties against numerical effects. All simulations are advanced until at least 15 sink particles have formed, and their properties are evaluated at this time ($t_{15}$). 

\subsection{The location of stars, relative to the ionising star}%

Fig. \ref{FIG:MSINK} shows how the formation radius, $R_{_{\rm FORM}}$ (i.e. the distance from the ionising star to a sink when it is first created), varies with the sink formation time, $t_{_{\rm FORM}}$. As ${\cal D}$ increases, the sinks form later and at larger radii. We derive a linear fit to the distribution of sinks in the ($R_{_{\rm FORM}},t_{_{\rm FORM}}$)-plane, using a $\chi^2$-minimization method. The best fit has a slope of $7.1\;{\rm km/s}$ (see dashed line in Fig. \ref{FIG:MSINK}). This velocity is comparable to the radial velocity of the ionisation front, and shows that stars are progressively triggered by the expansion of the H{\sc ii} region.

\subsection{The location of stars, relative to the ionisation front}%

\begin{figure}
\includegraphics[width=90mm]{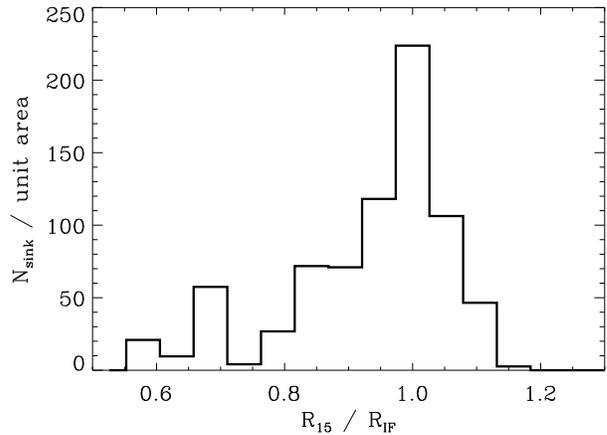} 
\includegraphics[width=90mm]{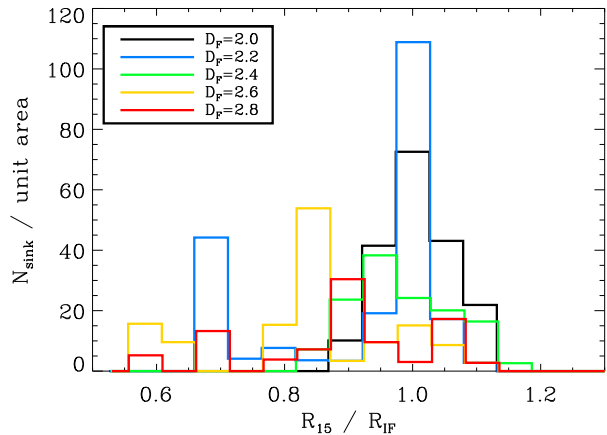} 
\caption{Histograms of the number of sinks at $t_{_{\rm 15}}$, as a function of the radial distance of the sink from the ionising star, $R_{_{\rm 15}}$, divided by the radial distance of the ionisation front from the ionising star, $R_{_{\rm IF}}$, in the same direction. The number counts are scaled by area of the annulus corresponding to each bin and thus represent a surface density. {\sc Top:} Total number count compiled using all sinks in all simulations. {\sc Bottom:} Number counts compiled for each fractal dimension.}
\label{FIG:RHIST}
\end{figure}

Fig. \ref{FIG:RHIST} shows two histograms of the number of sinks as a function of $R_{_{\rm 15}}/R_{_{\rm IF}}$, where $R_{_{\rm IF}}$ is the three-dimensional distance from the ionising star to the ionisation front along the line from the ionising star to the sink. The number of sinks is divided by the area of the annulus corresponding to each bin, so as to yield a surface density \citep[see][]{Thompson2012}. For simplicity, we are assuming that a 2-dimensional projection of the system along an arbitrary line of sight would on average yield the same ratio $R_{_{\rm 15}}/R_{_{\rm IF}}$. In the top panel, we compile all sinks formed in all simulations into one histogram. There is a clear over-density of triggered stars at, or close to, $R_{_{\rm IF}}$. In the bottom panel, we repeat the same analysis but distinguish the distributions for different $\mathcal{D}$. For low ${\cal D}$, the sinks typically stay ahead of the ionisation front, because the gas in the clumps from which these sinks are formed has been accelerated by the rocket effect. Therefore the sinks that condense out of it have significant radial velocities, whereas the expansion of the ionisation front is slowing down. Conversely, for high ${\cal D}$ the sinks form in the heads of pillars and are frequently left behind in the H{\sc ii} region, thus leading to a flat distribution of $R_{_{\rm 15}}/R_{_{\rm IF}} $.

Overall, the derived distribution of triggered stars compares remarkably well with the observational findings of \citet{Thompson2012}, who study the over-density of young stellar objects around {\it Spitzer} bubbles. However, they find a significantly enhanced source density at small radii ($R_{_{\rm 15}}/R_{_{\rm IF}} < 0.5$), which is not present in our analysis. One reason for this is the fact that they use projected positions of stars, whereas their estimate of the radius of the H{\sc ii} region is determined by its lateral extent. Thus, for a star on the far or near side of the H{\sc ii} region, close to the line of sight through the ionising star(s), $R_{_{\rm 15}}/R_{_{\rm IF}}$ will appear much smaller than the true three-dimensional value.


\begin{figure}
\includegraphics[width=92mm]{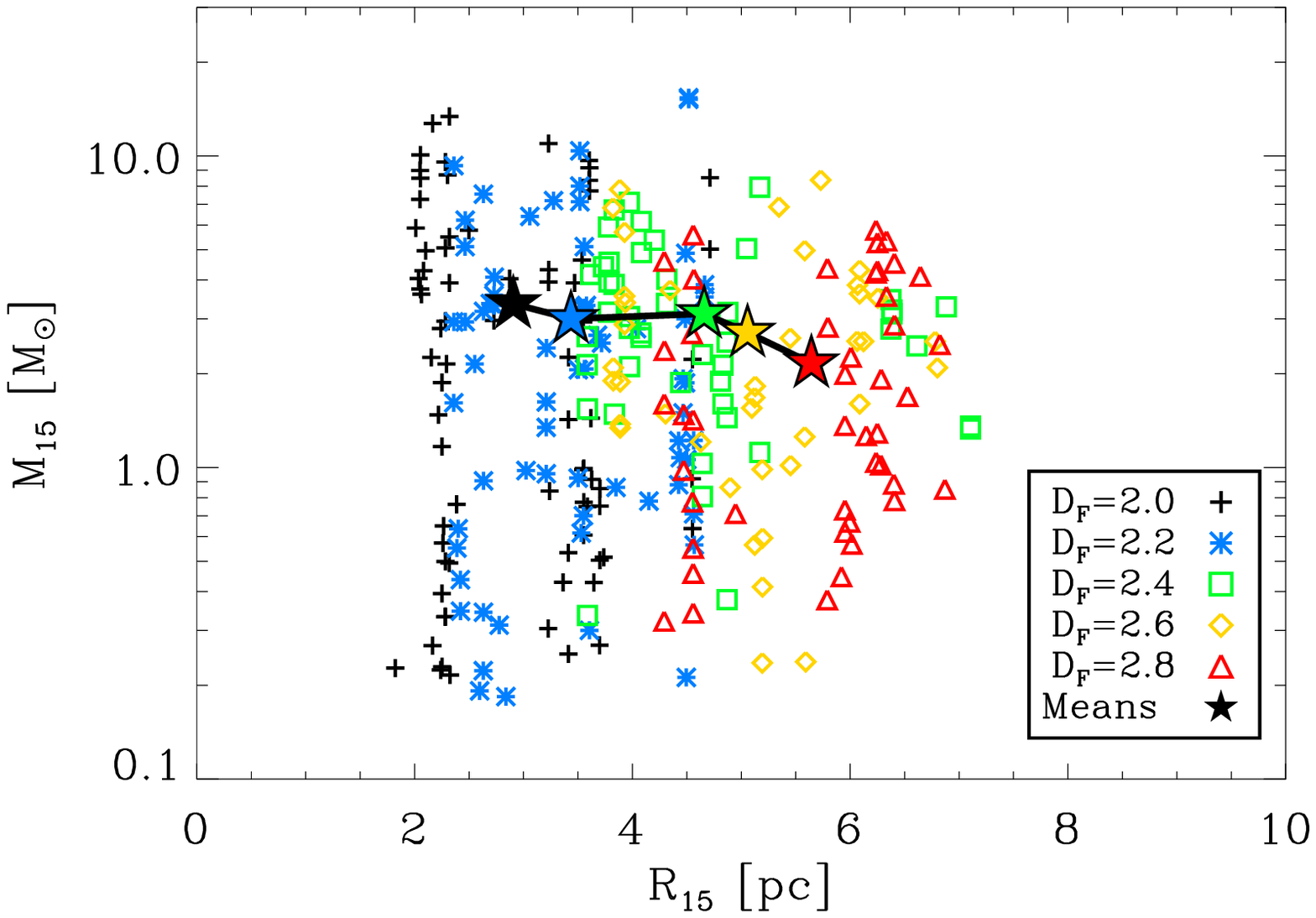} \\
\includegraphics[width=90mm]{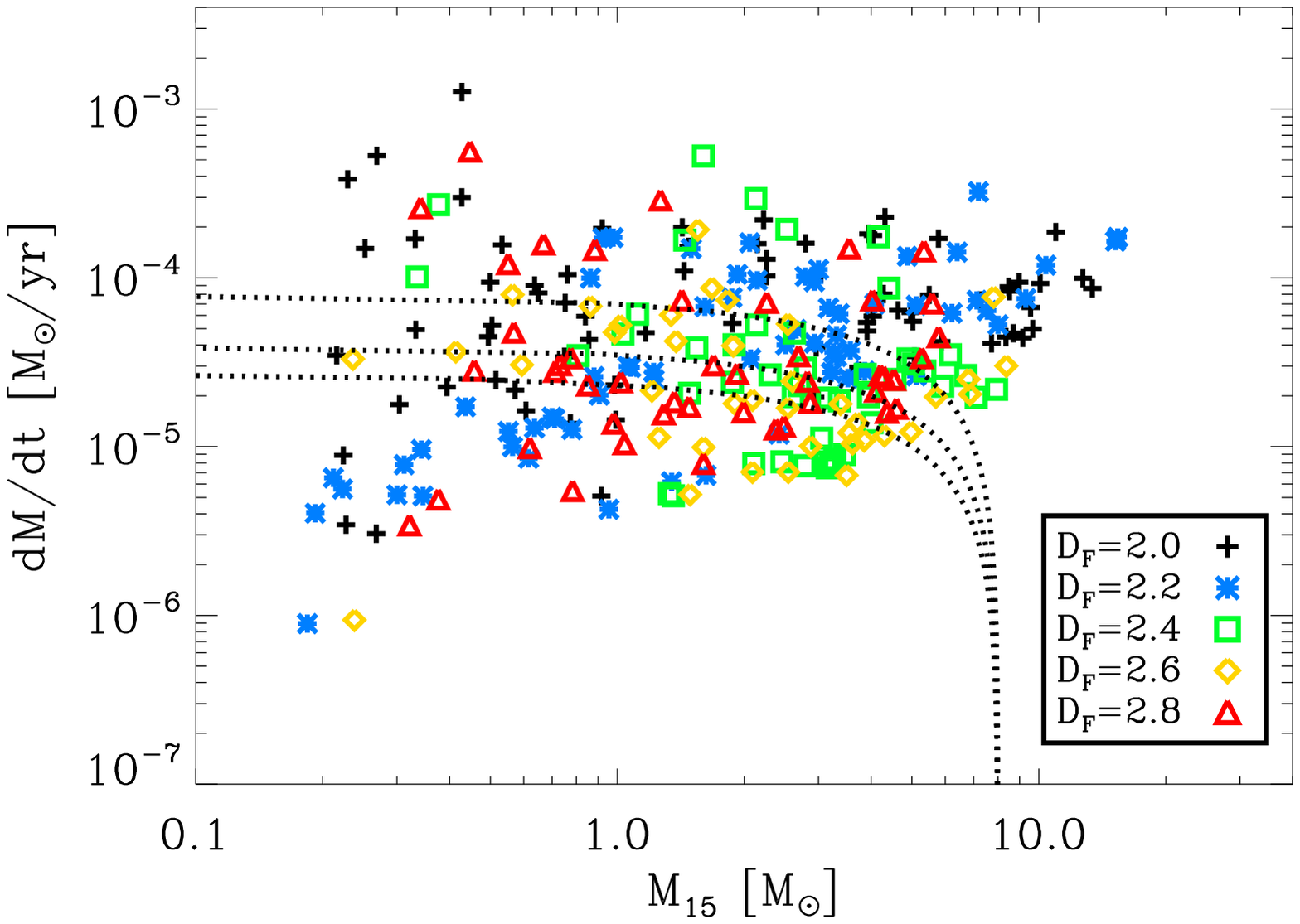}
\caption{{\sc Top panel:} Final sink mass, $M_{_{\rm 15}}$, as a function of final radial distance from the center, $R_{_{\rm 15}}$. Results obtained with different ${\cal D}$ are represented with different colours and different symbols. The star symbols show the mean for each value of ${\cal D}$. {\sc Bottom panel:} Mean mass accretion rate as a function of sink mass at $t_{_{\rm 15}}$. Sinks located above the dotted lines can evolve into a massive star with $M > 8\;{\rm M}_\odot$ if they continue accreting at their previous rate for 0.1 Myr (top line), 0.2 Myr (middle line), or 0.3 Myr (bottom line). }
\label{FIG:MASSES}
\end{figure}

\subsection{The masses of stars}%
Observational results from the Milky Way project \citep{Kendrew2012} and \citet{Deharveng2010} indicate that approximately 20\% of young, massive star formation may have been triggered. This estimate is also in agreement with \citet{Thompson2012}, who derive 14\% - 30\%.

Fig. \ref{FIG:MASSES} (top panel) shows how the sink mass, $M_{_{\rm 15}}$, varies with the three-dimensional radial distance from the center $R_{_{\rm 15}}$ at $t_{_{\rm 15}}$. Massive sinks ($M\!>\!8\,{\rm M}_{_\odot}$) are quite common for low ${\cal D}$ (shell-dominated morphology), and much rarer for high ${\cal D}$ (pillar-dominated morphology). The sinks seem to be aligned in vertical stripes, which are caused by two effects. First, $t_{15}$ and therefore the mean ionisation front radius varies for the three different realisations of every $\mathcal{D}$, which causes preferential triggering at different radii. Second, in some cases multiple arcs form around the H{\sc ii} region and therefore the sinks can be clustered about different radii. 

In the bottom panel of Fig. \ref{FIG:MASSES} we plot the mean mass accretion rates onto sinks up to $t_{_{\rm 15}}$, $\dot{M}$, as a function of their masses, $M_{_{\rm 15}}$. Above and to the right of the dotted lines a sink that continues to accrete at the observed rate for $0.1\,,0.2\;{\rm or}\;0.3\,{\rm Myr}$ will exceed $8\,{\rm M}_{_\odot}$, and therefore would be classified as a massive star. We see that for low ${\cal D}$ a significant fraction of sinks either are already, or will soon be, massive in this sense, whereas for higher ${\cal D}$ fewer of them are destined to be massive. The simulations with low-${\cal D}$ also appear to produce more low-mass stars, i.e. a bigger range of masses at both extremes is produced. The mass accretion rates are generally quite high ($\dot{M} \sim 5\times 10^{-5}\;{\rm M}_\odot/{\rm yr}$), which is not unexpected for these early stages of star formation. However, they decline as soon as the surrounding cold material has been accreted onto a sink, or ablated by the ionising radiation.

With respect to triggered star formation, the major limitation of the simulations presented here is that radiative and mechanical feedback from newly-formed stars is not included. Therefore, the quoted mass accretion rates are upper limits and we probably over-estimate the number of massive stars formed. At $\;t_{_{\rm 15}}$ the results are still credible, since the percentage of sinks with $M \ge 8\;{\rm M}_\odot$ is only 6.7\%. If all sinks were to continue accreting at their measured rate after $t_{_{\rm 15}}$, by $t_{_{\rm 15}} +0.1$ Myr the percentage of sinks with $M \ge 8\;{\rm M}_\odot$ would be $\sim$ 25\%. If feedback from newly-formed stars were included, it is not clear whether such a high percentage of massive stars would be able to form. 


\subsection{The clustering of stars}%

\begin{figure}
\includegraphics[width=90mm]{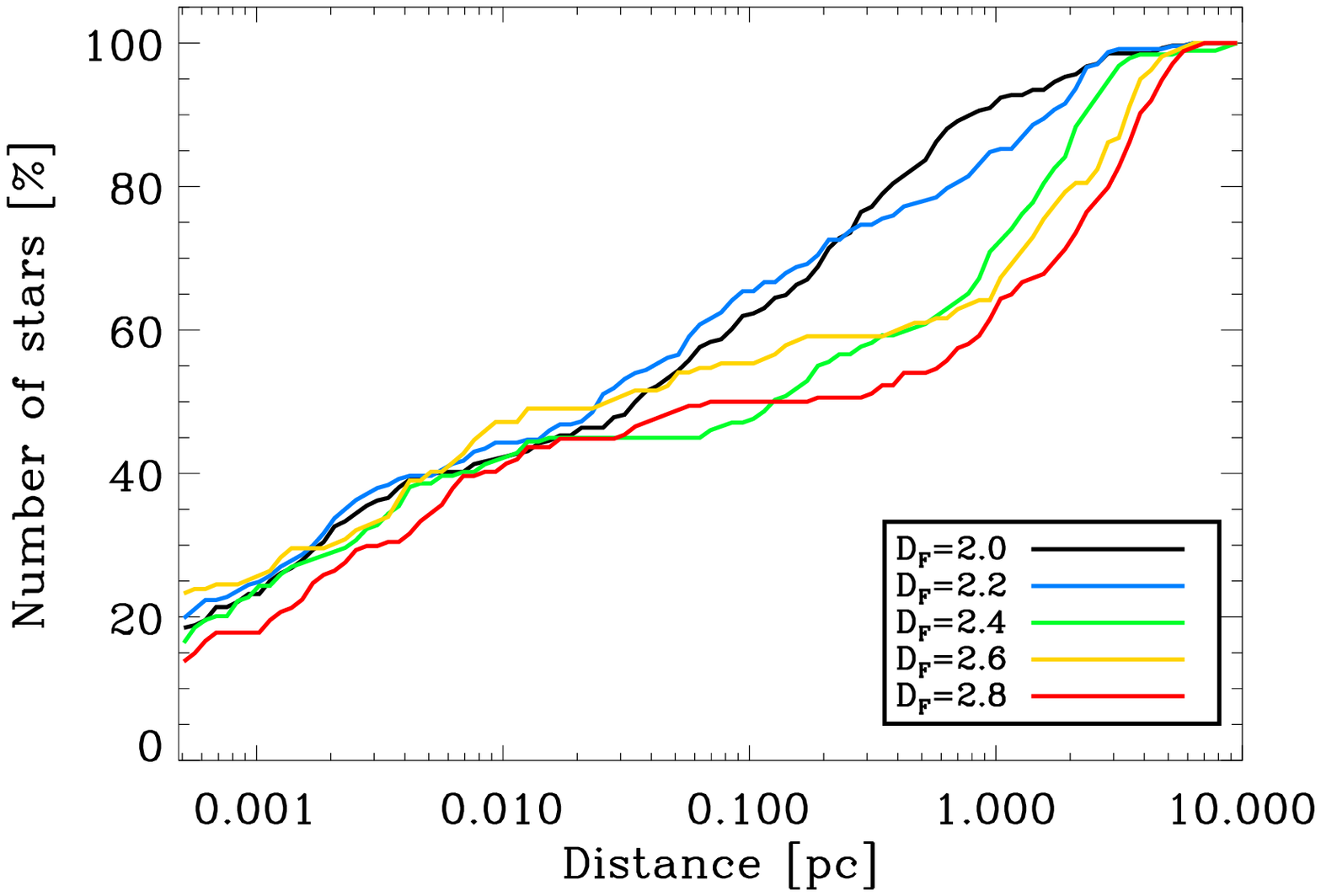} 
\includegraphics[width=90mm]{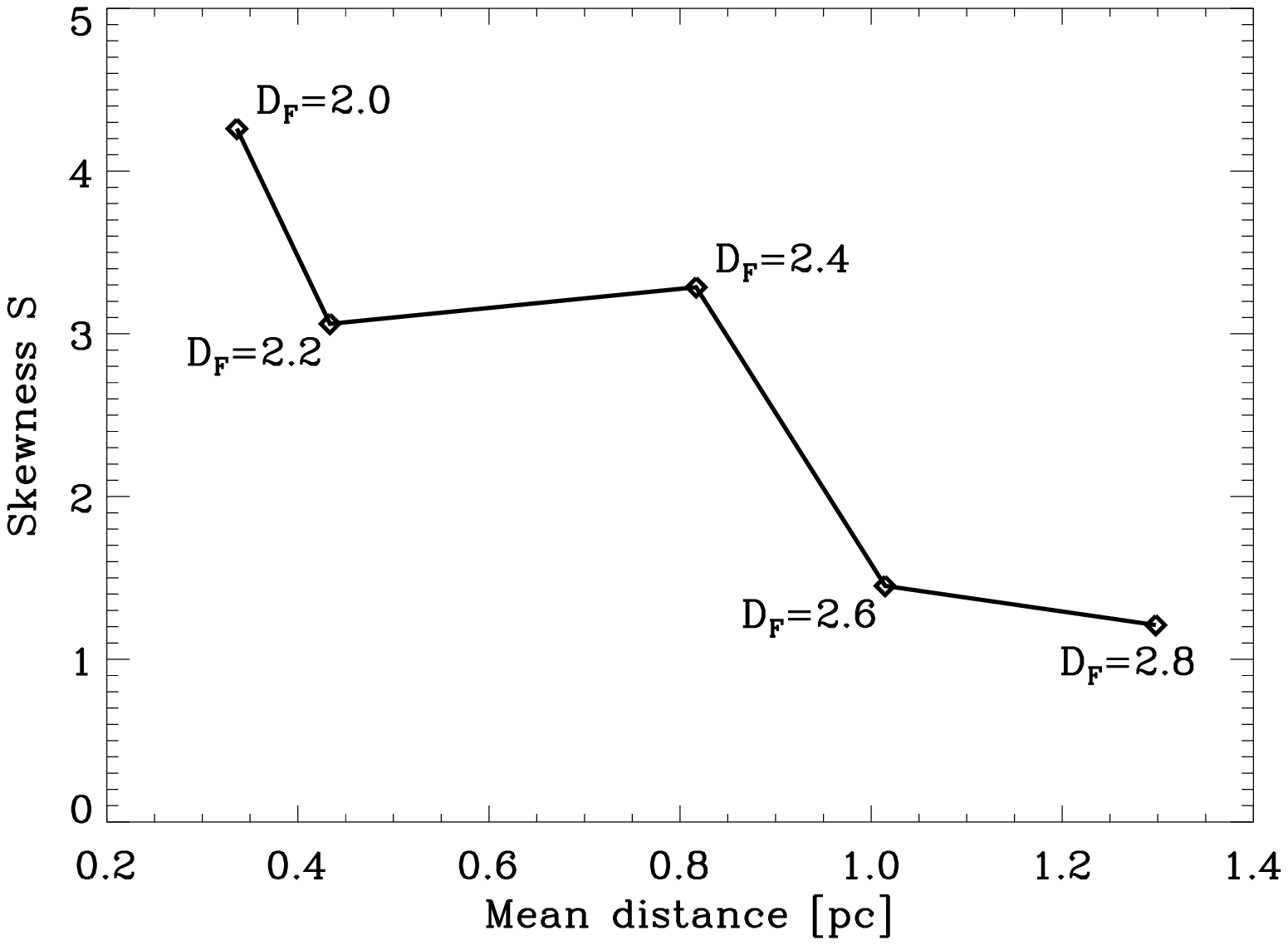} 
\caption{{\sc Top panel:} The cumulative distribution of the edge lengths derived from a Minimum Spanning Tree analysis of the sinks in each simulation at $t_{15}$. For each simulation, the MST is constructed using three projections, onto the fundamental Cartesian planes, and the results added to improve statistics.  {\sc Bottom panel:} The skew of the edge length distributions plotted against the mean separation. }
\label{FIG:MST}
\end{figure}

For each simulation, at $t_{15}$, we perform a Minimum Spanning Tree (MST) analysis. To construct an MST, we project all the star positions onto a plane, and then identify the system of straight lines (``edges") with minimum total length that links all the stars together; for an ensemble of ${\cal N}_{_\star}$ stars, there are ${\cal N}_{_\star}-1$ edges, and no closed loops. Having done this, we analyse the distribution of edge lengths, $\ell$. To improve the statistics, we project the star positions onto each of the fundamental Cartesian planes, and we consider all three realisations, so we end up with ${\cal N}_\ell=9({\cal N}_{_\star}-1)$ edge-lengths. If we define the $k$th moment about the mean, for the ensemble of edge-lengths,
\begin{eqnarray}
m_k&=&\overline{\left(\ell-\bar{\ell}\right)^k}\,,
\end{eqnarray}
the standard deviation of the ensemble is $\sigma_\ell=m_2^{1/2}$, and the skewness of the ensemble is $\gamma_\ell=m_3/\sigma_\ell^3$; the standard deviation is a measure of the width of the distribution, and the skewness is a measure of the asymmetry of the distribution. Fig. \ref{FIG:MST} shows the cumulative distribution of edge-lengths, for the different ${\cal D}$ values (top plot), and the skewness plotted against the mean (bottom plot). These plots demonstrate that with low ${\cal D}$ the stars are strongly clustered, whereas with high ${\cal D}$ they are more uniformly distributed. From the top plot we see that with ${\cal D}\!=\!2.0$, $\,\sim\! 90\%$ of edges are less than $1\,{\rm pc}$, whereas, with ${\cal D}\!=\!2.8$, only $\,\sim\! 62\%$ of edges are less than $1\,{\rm pc}$. From the bottom plot we see that with ${\cal D}\!=\!2.0$, the mean, $\mu_\ell$, is small, because most of the stars are in compact groups connected by small $\ell$, but the skew, $\gamma_\ell$ is large, because there is a significant tail of large $\ell$ that connect up the individual groups. As ${\cal D}$ is increased, $\mu_\ell$ increases because the stars become more homogeneously distributed, and their nearest neighbours are typically further away. The skewness tends to decrease, although not monotonically; it would be interesting to improve the statistics and explore whether this non-monotonicity is simply the result of small-number statistics.

\section{Conclusions}\label{SEC:CONC}%

In this paper, we use high-resolution 3D SPH simulations to explore the effect of a single O7 star emitting photons at $10^{49}\,{\rm s}^{-1}$ and located at the centre of a molecular cloud with mass $10^4\,{\rm M}_{_\odot}$ and radius $6.4\,{\rm pc}$. We focus on the statistics of dense clumps and triggered star formation, as a function of the initial fractal dimension,  ${\cal D}$, of the molecular cloud into which the H{\sc ii} region expands. We find that most properties show a clear correlation with ${\cal D}$.

Cold clumps form due to the sweeping up of gas by the H{\sc ii} region. The clumps are pushed outward by the rocket effect and grow in mass by collecting material in a snowplough manner. Thus, large clumps cover a bigger surface area and may accrete faster, even though the growth is not caused by self-gravity. 
\begin{itemize}
 \item For low ${\cal D} \le 2.2$ (shell-dominated regime), we find a small number of massive clumps, whereas high ${\cal D} \ge 2.6$ (pillar-dominated regime) results in many low-mass clumps. 
\item The clumps have trans- to super-sonic internal velocity dispersions. For non-star-forming clumps the internal velocity dispersion increases with clump size following Larson's Relation. For star-forming clumps the internal velocity dispersion is significantly higher than predicted by Larson's Relation. 
\item The resulting CMFs are well fitted by power-laws, with the slope increasing with increasing ${\cal D}$. Typically observed CMF slopes of $-0.7$ are recovered for intermediate ${\cal D}$.
\end{itemize}

The statistical properties of triggered stars are also well correlated with ${\cal D}$. On average, clouds with lower ${\cal D}$
\begin{itemize}
 \item form stars earlier in the simulation and at smaller distances from the ionising source (these stars are mostly located within the dense shell-like structures present for lower {\cal D}; for higher {\cal D} most stars sit in the tips of pillar-like structures);
 \item  are more prone to massive star formation; 
 \item  form mainly small star clusters, whereas for higher ${\cal D}$ star formation occurs in small-N multiple systems spaced at large distances from one another.
\end{itemize}

Stars are strongly concentrated near the ionisation front ($R_{_{\rm 15}}/R_{_{\rm IF}}=1$), but stars that form in pillars (high ${\cal D}$) tend to be left behind within the H{\sc ii} region.


\section*{Acknowledgments}%
We thank the anonymous referee for helpful comments and suggestions, which helped us to improve the paper. 
SKW thanks D. Kruijssen for useful and interesting discussions on the manuscript, and the Deutsche Forschungsgemeinschaft (DFG) for funding through the SPP 1573 'The physics of the interstellar medium'. SKW and AW further acknowledge the Marie Curie {\sc rtn constellation}. RW acknowledges the support of the Czech Science Foundation grant 209/12/1795 and by the project RVO:67985815. The simulations have been performed on the Cardiff {\sc arcca} Cluster. T.G.B. acknowledges support from STFC grant ST/J001511/1.

\bibliographystyle{mn2e}
\bibliography{references}%

\clearpage

\end{document}